# Strength, transformation toughening and fracture dynamics of rocksalt-structure $Ti_{1-x}Al_xN$ ($0 \leq x \leq 0.75$) alloys

D.G. Sangiovanni,[1,2*] F. Tasnádi,[1] L.J.S. Johnson,[3] M. Odén,[1] I.A. Abrikosov[1]

[1]Department of Physics, Chemistry and Biology (IFM) Linköping University,
SE-581 83, Linköping, Sweden

[2]ICAMS, Ruhr-Universität Bochum, D-44780 Bochum, Germany

[3]Sandvik Coromant, 126 80 Stockholm, Sweden

*Ab initio*-calculated ideal strength and toughness describe the upper limits for mechanical properties attainable in real systems and can, therefore, be used in selection criteria for materials design. We employ density-functional *ab initio* molecular dynamics (AIMD) to investigate the mechanical properties of defect-free rocksalt-structure (B1) TiN and B1 $Ti_{1-x}Al_xN$ (x = 0.25, 0.5, 0.75) solid solutions subject to [001], [110], and [111] tensile deformation at room temperature. We determine the alloys' ideal strength and toughness, elastic responses, and ability to plastically deform up to fracture as a function of the Al content. Overall, TiN exhibits greater ideal moduli of resilience and tensile strengths than (Ti,Al)N solid solutions. Nevertheless, AIMD modeling shows that, irrespective of the strain direction, the binary compound systematically fractures by brittle cleavage at its yield point. The simulations also indicate that $Ti_{0.5}Al_{0.5}N$ and $Ti_{0.25}Al_{0.75}N$ solid solutions are inherently more resistant to fracture and possess much greater toughness than TiN due to the activation of local structural transformations (primarily of B1 → wurtzite type) beyond the elastic-response regime. In sharp contrast, (Ti,Al)N alloys with 25% Al exhibit similar brittleness as TiN. The results of this work are examples of the limitations of elasticity-based criteria for prediction of strength, brittleness, ductility, and toughness in materials able to undergo phase transitions with loading. Comparing present and previous findings, we suggest a general principle for design of hard ceramic solid solutions which are thermodynamically inclined to dissipate extreme mechanical stresses via transformation toughening mechanisms.

*Corresponding author: davide.sangiovanni@liu.se

## I. Introduction

Hard, refractory rocksalt-structure (B1) titanium aluminum nitride [(Ti,Al)N] ceramics are extensively applied as wear and oxidation resistant protective coatings on cutting tools and engine components [1, 2]. The (Ti,Al)N parent binary phases – cubic rocksalt (B1) TiN and hexagonal wurtzite (B4) AlN – are immiscible at ambient conditions [3, 4]. Nevertheless, far-from-equilibrium synthesis methods as, e.g., vapor deposition techniques [5], allow the kinetic stabilization of single-phase B1 $Ti_{1-x}Al_xN$ over wide metal compositional ranges (up to x≈0.9) [6, 7]. During high-temperature operation (≈1000 – 1200 K), B1 (Ti,Al)N alloys undergo spinodal decomposition into strained, coherent B1 AlN-rich / B1 TiN-rich domains. This, in turn, greatly enhances the material's hardness thus improving the performance of the coating [8].

While single-phase materials generally become softer with temperature [9, 10], alloys as (Ti,Al)N are of considerable technological importance due to the spinodally-induced age hardening effect. Over the past decades, several studies [8, 11-18] focused on understanding the surface reactivity and thermodynamics of phase segregation in order to design (Ti,Al)N-based coatings with superior thermal stability and hinder B1→B4 AlN-domain transformations [19-21]. In contrast, the toughness and resistance to fracture of (Ti,Al)N and (Ti,Al)N-based solid solutions have not been investigated as extensively, with a few studies available in the literature [22-25]. Recent experiments suggest that, although detrimental for the alloy hardness, the nucleation of wurtzite phases in B1 AlN-rich regions does not affect, or is even beneficial for, the coating toughness by inhibiting crack formation and/or propagation [26, 27]. Nonetheless, the presence of grain boundaries and voids, which act as *weakest links* [28] in polycrystalline samples, ultimately controls the resistance to fracture of (Ti,Al)N, thus preventing the possibility of describing the alloy mechanical response as a function of metal composition. Moreover, the fact that (Ti,Al)N ceramics are typically synthesized in the form of thin films complicates the experimental evaluation of their strength and toughness. These problems render first-principles approaches an indispensable tool for the investigation of the mechanical properties of single-crystal B1 (Ti,Al)N solid solutions.

As a first step toward understanding the intrinsic ability of defect-free B1 Ti$_{1-x}$Al$_x$N to withstand loading and plastically deform, we employ *ab initio* molecular dynamics (AIMD) simulations at 300 K – temperature at which refractory ceramics are typically brittle [29-31] – to investigate the effects induced by an increasing Al content (x=0, 0.25, 0.5, 0.75) on the alloys' responses to [001], [110], and [111] tensile deformation [32]. The simulations allow us to observe the dynamics of brittle cleavage vs. lattice-transformation-induced toughening as a function of the metal composition.

## II. Computational methods

AIMD [33] simulations are performed using VASP [34-36] implemented with the projector augmented wave method [37]. The electronic exchange and correlation energies are parameterized according to the generalized gradient approximation of Perdew, Burke, and Ernzerhof [38]. All AIMD simulations employ Γ-point sampling of the reciprocal space and planewave cutoff energies of 300 eV. The nuclear equations of motion are integrated at 1 fs timesteps, using an energy convergence criterion of 10$_{-5}$ eV/supercell for the ionic iterations. Prior to modeling tensile deformation, the supercell structural parameters are evaluated via NPT sampling of the configurational space (Parrinello-Rahman barostat [39] and Langevin thermostat set to 300 K). Subsequently, AIMD within the NVT ensemble (Nose-Hoover thermostat, with a Nose mass of 40 fs) is used to equilibrate the structures at 300 K during three additional ps, ensuring that the time-averaged stress components |σ$_{xx}$|, |σ$_{yy}$|, and |σ$_{zz}$| are ≤ 0.3 GPa.

In order to model tensile deformation, as well as shear deformation leading to lattice slip (results presented in a parallel study [40]), the Ti$_{1-x}$Al$_x$N supercells are conveniently oriented with their *z* vertical axis along [001]-, [110]-, and [111] directions, and with lateral x axes along the [1–10] Burgers vector direction (**Fig. 1**). [*h k l*]-oriented supercells are denoted below as Ti$_{1-x}$Al$_x$N(*h k l*), where *h*, *k*, and *l* are Miller indexes. B1 Ti$_{1-x}$Al$_x$N (0 ≤ x ≤ 0.75) simulation boxes contain 288 metal and 288 nitrogen atoms (576 ideal B1 sites with 24 atomic layers orthogonal to the tensile strain $z$ direction), applying periodic boundary conditions in three dimensions (**Fig. 1**). Al and Ti atoms are stochastically arranged on the cation sublattice, thus ensuring negligible degrees of short-range metal ordering. Tensile deformation is carried out by following the scheme detailed in Ref. [32]. Briefly, at each strain step (2% of the supercell length along $z$), the structures are (i) first rapidly equilibrated by isokinetic velocity-rescaling during 300 fs and (ii) then maintained at the same temperature during additional 2.7 ps using the Nose-Hoover thermostat. At each strain step, tensile $\sigma_{zz}$ stresses are determined by averaging $\sigma_{zz}$ stresses calculated for the 500 final AIMD configuration. Moduli of ideal tensile resilience $U_R$ – energy density accumulated during elastic deformation (i.e., up to the yield point) – and ideal tensile toughness $U_T$ – energy density absorbed up to fracture – are calculated by integrating the area underlying stress vs. strain curves up to the yield $\delta_y$ and fracture $\delta_f$ strains, respectively. The supercell size along the lateral $x$ and $y$ directions is maintained unvaried during tensile deformation. Images and videos are generated using the visual molecular dynamics [41] software.

## III. Results and discussion

**Figure 2** illustrates the dependence of $\sigma_{zz}$ stresses vs. uniaxial elongation of B1 Ti$_{1-x}$Al$_x$N solid solutions determined via AIMD simulations at room temperature. The slopes of stress vs. strain curves [42] within the alloy elastic-response up to $\delta=4\%$ are used to calculate (see equations 2 and 4 in Ref. [43]) the $C_{11}$ and $C_{12}$ elastic constants as a function of $x$. AIMD results yield $C_{11}$ elastic stiffnesses which, for x increasing from 0 to 0.75, monotonically decrease from 650±50 GPa to 528±38 GPa (**Table I**). Noting that $x$ and $y$ supercell axes are parallel to <110> crystallographic directions (**Fig. 1(a)**), the $C_{12}$ elastic constant can be evaluated via 45° rotation of the stress tensor within the $xy$ plane. The calculated $C_{12}$ values monotonically increase with the Al content from 128±6 GPa (for x = 0) to 174±8 GPa (for x = 0.75). Accordingly, the bulk moduli $B$ remain approximately constant, or exhibit slight reductions with Al substitutions (**Table I**). The uncertainties on the $C_{11}$ and $C_{12}$ values arise from the sensitivity of calculated elastic constants on the choice of strain ranges and deformation tensors [44] and the presence of small residual stress components in the relaxed supercell structures. The influence of metal-species arrangements, which produces a large scatter on $C_{11}$ and $C_{12}$ values calculated for anharmonic transition-metal nitride alloys [45], is expected to have negligible effects on the elastic response of TiN and (Ti,Al)N solid solutions. The trends in, and absolute values of $C_{11}$, $C_{12}$, and $B$ vs. x (**Table I**) agree, within uncertainty ranges, with those reported by previous *ab initio* calculations at 0 K [46, 47] and AIMD simulations at room temperature [43, 48].

**Fig. 2(a)** and **Table II** show that the ideal $Ti_{1-x}Al_xN(001)$ tensile strength $\gamma_{[001]}$ – the vertical $\sigma_{zz}$ maximum stress obtained at the yield point during [001] elongation – remains approximately constant at 39 GPa for Al contents in the range $0 \leq x \leq 0.5$. An increase in Al metal content to 75% induces a slight $\gamma_{[001]}$ reduction to 37 GPa. The yield points of TiN(001) and $Ti_{0.75}Al_{0.25}N(001)$ are reached at 10% elongation, while slightly larger values (12%) are obtained for $Ti_{0.5}Al_{0.5}N(001)$ and $Ti_{0.25}Al_{0.75}N(001)$. Conversely to the trend observed for the $C_{11}$ elastic constants, which demonstrates a reduction in [001] stiffness for increasing x (**Fig. 2(a)** and **Table I**), the alloys with high Al contents display larger moduli of resilience ($U_{R[001]}$ = 3.0 and 2.8 GPa) than TiN ($U_{R[001]}$ = 2.5 GPa) and $Ti_{0.75}Al_{0.25}N$ ($U_{R[001]}$ = 2.4 GPa), see **Table II**. To summarize, AIMD simulations demonstrate that the room-temperature $Ti_{1-x}Al_xN$ mechanical response to [001] tensile deformation up to yield points, which approximate the limit for the elastic response, is not dramatically affected by Al substitutions. This is consistent with the fact that covalent N (p) – metal (d-$e_g$) bonding states remain fully occupied even though the valence electron concentration of B1 $Ti_{1-x}Al_xN$ solid solutions decreases from 9 e–/f.u. (for TiN) to 8.25 e–/f.u. (for $Ti_{0.25}Al_{0.75}N$) [49-51]. Nonetheless, simulation results (see below) suggest that an increasing Al content significantly promotes the alloys' ability to plastically deform, thus improving the material's toughness.

In agreement with AIMD results of Ref. [32, 52], an extension of TiN(001) beyond its tensile yield point (≈10%) leads to brittle fracture of the material. AIMD modeling reveals that 25% replacement of Ti atoms with Al induces negligible effects on the alloy plastic response to [001] uniaxial deformation; cubic Ti$_{0.75}$Al$_{0.25}$N(001) solid solutions remain brittle and undergo sudden cleavage on the (001) plane at strains larger than 10% (see **Fig. 2(a)**, **Fig. 3** and **Table II**). In sharp contrast, Ti$_{1-x}$Al$_x$N alloys with Al contents x ≥ 0.5 are considerably more resistant to fracture than TiN and Ti$_{0.75}$Al$_{0.25}$N. This is due to their ability to undergo local structural changes into wurtzite-like atomic environments when the elongation overcomes their yield points (see **Figs. 2(a)**, **4**, and **5**).

The modifications in the bonding network that become operative in B1 Ti$_{0.5}$Al$_{0.5}$N(001) and Ti$_{0.25}$Al$_{0.75}$N(001) solid solutions at high tensile strains can be rationalized on the basis of transformation pathways induced by pressure in wurtzite group-III nitrides, such as AlN (see examples of strain-mediated B4→B1 AlN transitions in Ref. [53]), which is a border (x=1) case for the investigated Ti$_{1-x}$Al$_x$N system [54, 55]. Tetragonal [56] and hexagonal (graphitic-like, boron-nitride prototype, B$_k$) crystal structures are the predicted transition states along the B4 → B1 transformation path of group-III nitrides and other semiconductors (see, e.g., figure 1 in Ref. [57]). The B4 → B1 AlN transformation path energetically favors the B$_k$ intermediate state: compression of the wurtzite lattice along the [0001] direction followed by shear deformation within the (0001) B$_k$ plane [55]. It is therefore expected that the inverse (B1 → B4) AlN phase transition should also preferentially occur through the B$_k$ metastable configuration.

At deviance with the transformation path predicted for AlN, AIMD simulations show that B1 Ti$_{0.5}$Al$_{0.5}$N(001) and B1 Ti$_{0.25}$Al$_{0.75}$N(001) elongated beyond their yield point exhibit buckling of (001) atomic planes, which correspond to a tetragonal state (see schematic illustration in **Fig. 6**). The formation of tetragonal (Ti,Al)N domains that precede the appearance of wurtzite-like environments is reminiscent of the solid→solid transformation path predicted for other B4-structure crystals as, e.g., GaN and ZnO [56]. As indicated in Ref. [56], B4→tetragonal→B1 transitions are presumably favored due to the presence of d-electrons (note that B1 Ti$_{1-x}$Al$_x$N with 0 ≤ x ≤ 0.75 is a conductor with d-states at the Fermi level [58]). The lattice transformation active in Ti$_{0.5}$Al$_{0.5}$N and Ti$_{0.25}$Al$_{0.75}$N ultimately results in a considerably enhanced resistance to fracture and a substantially increased toughness (area underlying stress/strain curves) during [001] tensile deformation (see **Table II**).

Given that (001) surfaces in B1-structure ceramics have much lower formation energies than (110) and (111) terminations [59] – i.e., crack formation is energetically favored on (001) planes – the AIMD results discussed above for Ti$_{1-x}$Al$_x$N(001) tensile elongation are of major importance for assessing the (Ti,Al)N resistance to fracture as a function of the Al content. Nevertheless, complementary AIMD results obtained for crystals strained along [110] and [111] directions (see below) provide a more comprehensive understanding for the effects induced by Al on the inherent mechanical response of B1 Ti$_{1-x}$Al$_x$N to elongation.

For an increasing Al concentration, tensile-strained Ti$_{1-x}$Al$_x$N(110) and Ti$_{1-x}$Al$_x$N(111) exhibit a monotonic increase in elastic stiffness (i.e., initial slope in $\sigma_{zz}$ vs. strain), accompanied by an overall reduction in U$_R$, (see **Fig. 2(b,c)** and **Table II**). These trends are opposite to that calculated for Ti$_{1-x}$Al$_x$N(001). In contrast, the $\gamma_{[110]}$ and $\gamma_{[111]}$ tensile strengths of the alloy are not significantly affected by Al substitutions. In fact, for each investigated strain direction, the relative strength variation with x remains within 10% (**Fig. 2** and **Table II**). Irrespective of the metal composition, AIMD results show that the relationship between alloy tensile strengths is $\gamma_{[111]}$ (64–71 GPa) > $\gamma_{[110]}$ (54–56 GPa) > $\gamma_{[001]}$ (37–39 GPa). This is consistent with the trend in surface formation energies E$_{s(111)}$ > E$_{s(110)}$ > E$_{s(001)}$ reported for B1-structure materials [59], that is, the uniaxial strength is related to the energy required to cleave the crystal on a plane normal to the elongation direction.

Although alloys with Al concentrations ≥50% present smaller U$_{R[110]}$ and U$_{R[111]}$ moduli of resilience than TiN and Ti$_{0.75}$Al$_{0.25}$N, as described below, AIMD simulations reveal that high Al contents are beneficial for the total tensile toughness U$_{T[110]}$ and U$_{T[111]}$ (see **Fig. 2(b,c)** and **Table II**). Combined with the results described above for [001]-strained materials, these findings indicate that the room-temperature mechanical properties of B1 Ti$_{1-x}$Al$_x$N are considerably improved by Al substitutions of ≥ 50%.

Consistent with AIMD results reported in a previous study [32, 52], TiN(110) undergoes sudden brittle failure when the [110] uniaxial strain reaches ≈18%, (**Fig. 2(b)**). The mechanical response of Ti$_{0.75}$Al$_{0.25}$N solid solutions to [110] elongation is nearly equivalent to that determined for the binary compound (**Fig. 2(b)**). AIMD simulation snapshots of Ti$_{0.75}$Al$_{0.25}$N(110) at a constant tensile strain of 18% display rapid (within 1.3 ps) bond snapping that causes brittle cleavage of the alloy (**Fig. 7**). Note that the fractured region follows a zig-zag pattern on (001) crystallographic planes. Conversely, B1 solid solutions that contain 50% and 75% Al undergo, after the yield point, local changes in bonding geometries which prevent sudden mechanical failure (in comparison to TiN(110) and Ti$_{0.75}$Al$_{0.25}$N(110)). The transformation toughening effect induced by Al substitutions in [110]-strained Ti$_{0.5}$Al$_{0.5}$N and Ti$_{0.25}$Al$_{0.75}$N is illustrated by AIMD snapshots in **Fig. 8** and **Fig. 9**, respectively.

As shown in **Fig. 2(b)**, vertical σ$_{zz}$ stresses meet the ideal Ti$_{0.5}$Al$_{0.5}$N(110) and Ti$_{0.25}$Al$_{0.75}$N(110) tensile strengths for an elongation of 14%. Up to that strain, both alloys maintain ideal octahedral atomic coordination (see upper-left panels in **Figs. 8** and **9**). A [110] deformation of 16%, activates local modifications in the bonding network (**Figs. 8** and **9**), as reflected by a drop in σ$_{zz}$ vs. strain in **Fig. 2(b)**. Thus, the mechanical response beyond the yield points of Ti$_{0.5}$Al$_{0.5}$N(110) and Ti$_{0.25}$Al$_{0.75}$N(110) solid solutions is dramatically different to those observed for TiN(110) [32] and Ti$_{0.75}$Al$_{0.25}$N(110). Fracture in Ti$_{0.5}$Al$_{0.5}$N(110) and Ti$_{0.25}$Al$_{0.75}$N(110) occurs in a more controlled manner. Relatively slow bond fraying accompanied by progressive void opening delays mechanical failure. In this regard, we should underline that, due to transformation toughening processes, the fracture points of Ti$_{0.5}$Al$_{0.5}$N(110) and Ti$_{0.25}$Al$_{0.75}$N(110) are not unambiguously identifiable. The AIMD bonding configurations hold the materials together up to 20% elongation, while rupture is identified by the appearance of several voids at ≈22–24% strain, see **Figs. 8** and **9**. More important, however, our results qualitatively demonstrate that relatively high Al contents in (Ti,Al)N lead to a superior resistance to fracture.

Similarly to the results obtained for B1 Ti$_{1-x}$Al$_x$N subject to [001] and [110] tensile strain, AIMD simulations of supercells deformed along [111] directions confirm that Al substitutions are beneficial for the alloy resistance to fracture. TiN(111) displays the highest γ$_{[111]}$ tensile strength (71 GPa) and resilience U$_{R[111]}$ = 7.3 GPa. As expected, the binary compound fractures in a brittle manner beyond its yield point (**Figs. 2(c)** and **10**). A constant elongation of 20% leads, within 1.8 ps, to bond snapping and crack opening primarily along (001) crystallographic planes. Ti$_{0.75}$Al$_{0.25}$N(111) displays a mechanical response to [111] deformation qualitatively similar to that of the binary nitride, that is, brittle fracture occurs within few ps at constant strain of 18%, **Fig. 2(c)**. Ti$_{0.5}$Al$_{0.5}$N(111) and Ti$_{0.25}$Al$_{0.75}$N(111) solid solutions reach their yield points at 14% strain, **Fig. 2(c)**, with all atoms maintaining octahedral coordination (upper-left panel in **Fig. 11**). As anticipated by the results determined for Ti$_{0.5}$Al$_{0.5}$N and Ti$_{0.25}$Al$_{0.75}$N alloys strained along [001] and [110] directions, a [111] deformation beyond the yield point activates local structural transformations which allow stress dissipation and prevent brittle fracture. AIMD snapshots in **Figs. 11** and **12** demonstrate that Ti$_{0.5}$Al$_{0.5}$N and Ti$_{0.25}$Al$_{0.75}$N break via a progressive, yet slow, reduction in bond densities induced by an increasing strain. A qualitative comparison with TiN (**Fig. 10**), reveals that elongations of ≈20% (**Fig. 11**) and ≈24% (**Fig. 12**) are necessary to completely open the crack in Ti$_{0.5}$Al$_{0.5}$N and Ti$_{0.25}$Al$_{0.75}$N, respectively. Overall, TiN(111) presents greater toughness than Ti$_{0.5}$Al$_{0.5}$N(111) (**Table II** and **Fig. 2(c)**). This is due to the fact that, while both materials fracture at 20% strain, the binary compound reaches mechanical yielding at a much higher elongation than the ternary alloy. In contrast, Ti$_{0.25}$Al$_{0.75}$N(111) solid solutions exhibit equal strength, but higher toughness, than TiN(111) owing to slow bond fraying which delays fracture up to an elongation of 22–24% (**Fig. 12**).

The transformation toughening effect observed via [110] and [111] elongation of B1 Ti$_{0.5}$Al$_{0.5}$N and Ti$_{0.25}$Al$_{0.75}$N is less pronounced than the mechanism induced by [001] strain because these deformation paths offer lower flexibility toward B1→tetragonal→B4 transitions (**Fig. 6**). Indeed, the bonding geometries visible in plastically-deformed domains of Ti$_{0.5}$Al$_{0.5}$N(110) (**Fig. 8**),

Ti$_{0.25}$Al$_{0.75}$N(110) (**Fig. 9**), Ti$_{0.5}$Al$_{0.5}$N(111) (**Fig. 11**), and Ti$_{0.25}$Al$_{0.75}$N(111) (**Fig. 12**) suggest that local structural amorphization takes place in response to extreme external stresses. Notably, this characteristic is not observable in TiN and Ti$_{0.75}$Al$_{0.25}$N, likely due to high stability of octahedral bonding configurations (see figure 7(c,d) in [32]) along [001], [110], and [111] uniaxial transformation paths.

Nanoindentation mechanical testing of single-crystal B1 TiN films demonstrates its inherently brittle nature [30]. On the other hand, experimental information for the mechanical properties of monolithic B1 Ti$_{1-x}$Al$_x$N solid solutions with $x \geq 0.5$ are not currently available. Nevertheless, the AIMD predictions of this work are qualitatively supported by the experimental observations of Bartosik et al. [27], which indicated that the resistance to fracture of single-phase B1 nanocrystalline Ti$_{0.4}$Al$_{0.6}$N solid solutions benefits from the formation of hexagonal B4 domains upon loading. Other experimental investigations also suggest that dual-phase wurtzite/cubic Ti$_{1-x}$Al$_x$N ($x \approx 0.75$) films possess high hardness (30 GPa) [60], which indirectly contributes to enhance the materials' toughness. Moreover, in B1-ZrN/B1-ZrAlN [61] and B1-CrN/B1-AlN [62] superlattices, stress-induced B1→B4 transformation in B1 AlN-rich domains has been demonstrated to significantly increase the materials' toughness.

The results of this work provide fundamental insights of the mechanical properties of B1 (Ti,Al)N solid solution ceramics during use. However, it is important to underline that the macroscopic mechanical behavior and resistance to fracture of polycrystalline B1 (Ti,Al)N coatings are primarily controlled by microstructural features such as grain size, texture, and grain boundary properties. For example, cracks can more easily initiate and propagate at the interfaces between crystallites where the density is lower and voids may be present. Nonetheless, the toughening mechanisms observed in AIMD simulations can operate within (Ti,Al)N grains of sufficiently large size (less affected by grain boundary properties), when tensile stresses build up inside the grain. The elongations at fracture $\delta_f$ shown in **Fig. 2** and **Table II** are indicative of the relatively ability of B1 (Ti,Al)N alloys with different metal compositions to endure deformation by undergoing local (nm length-scale) modifications in the bonding network.

It should also be emphasized that our present AIMD simulations pertain the mechanical behavior of (Ti,Al)N solid solutions at 300 K. At this temperature, spinodal decomposition is *kinetically blocked*, i.e., the temperature is not high enough to activate diffusion of vacancies (vacancy migration in B1 (Ti,Al)N systems requires energies in the range ≈2.5 – 4.5 eV [63-66]). In general, if the operation temperature of (Ti,Al)N coatings remains below ≈1000 K – typically the onset for decomposition – we find it unlikely that the spinodal decomposition process may occur faster than the strain-mediated lattice transformations seen here.

At a fundamental electronic-structure level, a relatively high Al metal content (≈60%) is expected to maximize the hardness of B1 (Ti,Al)N alloys. The effect stems from the fact that ≈8.4 $e_-$/f.u. in B1-structure transition-metal (carbo)nitride solid solutions fully populate strong *p-d* metal/N bonding states while leaving shear-sensitive *d-d* metallic states empty [49]. In contrast, a low occupancy of *d* states is detrimental for the ability of (Ti,Al)N to form metallic bonds upon shearing. That, in turn, has been suggested as a possible cause of brittleness [51]. Consistent with the analysis of Ref. [51], phenomenological ductility/brittleness predictions based on elastic constant values would also (erroneously) indicate that Al substitutions degrade the (Ti,Al)N resistance to fracture. For example, according to the criterion proposed by Pettifor [67], the decrease in $C_{12}$ - $C_{44}$ Cauchy's pressure suggests that B1 $Ti_{1-x}Al_xN$ solid solutions become progressively more brittle for increasing x (see figure 1 in Ref. [46]). However, DFT predictions of toughness vs. brittleness in (Ti,Al)N [51], primarily based on the analyses of the alloy elastic deformation, are unsuited to reveal the occurrence of transformation toughening mechanisms in the plastic regime.

B1 (Ti,Al)N ceramics are of enormous technological importance due to age-hardening induced by spinodal decomposition at elevated temperatures [68]. However, while the spinodal mechanism is kinetically-blocked at ambient conditions, DFT calculations at 0 K show that an Al metal content larger than ≈0.7 renders Ti$_{1-x}$Al$_x$N solid solutions energetically more stable in the wurtzite than in the rocksalt structure (see figure 3a in [69]). The results of present AIMD simulations, combined with those of Ref. [69], evidence a correlation between the phase stability of the alloys and their inherent room-temperature toughness vs. brittleness.

The mechanical behavior predicted by AIMD for TiN and Ti$_{0.75}$Al$_{0.25}$N (**Figs. 3, 7, 10,** and Ref. [32]) indicates that an Al content much lower than 0.5 causes B1 Ti$_{1-x}$Al$_x$N embrittlement. Presumably, the energy required to induce cleavage in these two systems is smaller than the one necessary to activate any local lattice transformation during uniaxial strain. In contrast, our results suggest that tuning the Ti$_{1-x}$Al$_x$N metal composition around the threshold value x≈0.7 [69] can be used to optimize the combination of strength and toughness of B1-structure alloys. Indeed, the relatively small E$_{B1}$-E$_{B4}$ energy difference calculated for Ti$_{0.5}$Al$_{0.5}$N [69] enables B1→B4-like transformations during [001] tensile deformation (**Figs. 3** and **5**), thus dissipating accumulated stresses and enhancing the material resistance to fracture. On the other hand, Ti$_{0.25}$Al$_{0.75}$N solid solutions (which can be synthesized as B1 single-phase films [6, 7]) would favorably crystallize in the B4 polymorph structure at ambient conditions [69]. Accordingly, Ti$_{0.25}$Al$_{0.75}$N is thermodynamically more inclined than Ti$_{0.5}$Al$_{0.5}$N to activate B1→B4 transformations under load. Consistent with this observation, AIMD shows that, for elongations progressively increasing beyond the yield point of the material, wurtzite-like domains grow faster in Ti$_{0.25}$Al$_{0.75}$N(001) than in Ti$_{0.5}$Al$_{0.5}$N(001). This is reflected by a greater drop in the stress of Ti$_{0.25}$Al$_{0.75}$N(001) visible between 12% and 14% strain (**Fig. 2(a)**).

That metastability is beneficial to enhance the mechanical performance of ceramics is not a new concept. For example, it has been shown that tuning the electron concentration to values near 9.5 e$_-$/f.u. sets hexagonal and cubic polymorph structures of transition-metal carbonitrides to similar energies. This, in turn, promotes formation of hexagonal stacking faults in cubic alloys, thus increasing hardness by hindering dislocation motion across the faults [70, 71]. Similarly, plastic deformation along 111 faults in B1 refractory carbonitrides can be assisted by providing facile deformation paths: the energy barrier of {111}<1–10> slip is reduced by synchro-shear mechanisms in B1 Ti$_{0.5}$W$_{0.5}$N solid solutions and B1-TiN/B1-WN$_x$ superlattices due to the preference of B1 WN-rich domains to transform in more stable hexagonal WC-structures [72, 73]. Analogous to the experimental findings for multilayer films of Yalamanchili et al. [61] and Schlögl et al. [62], in this work we show that alloying (in ideal defect-free structures) transition-metal nitrides with AlN can – beside spinodal age-hardening at elevated temperatures – enable B1→B4 transformation toughening mechanisms at 300 K, i.e., much lower than the typical brittle-to-ductile transition temperatures of refractory ceramics [29, 31].

## IV. Conclusions

AIMD simulations at 300 K are used to determine the inherent tensile strength, toughness, and resistance to fracture of defect-free B1 Ti$_{1-x}$Al$_x$N solid solutions ($0 \leq x \leq 0.75$). The results show that TiN and Ti$_{0.75}$Al$_{0.25}$N are strong materials, but cleave at their yield point via sudden bond snapping. In contrast, Ti$_{0.5}$Al$_{0.5}$N and Ti$_{0.25}$Al$_{0.75}$N exhibit similar strength, but significantly higher toughness than TiN and Ti$_{0.75}$Al$_{0.25}$N, due to the activation of local lattice transformations in the plastic-response regime which dissipate stress, thus preventing brittle failure. Overall, B1 Ti$_{0.25}$Al$_{0.75}$N solid solutions exhibit the best combination of room-temperature strength and toughness, due to an energetic preference toward the more stable B4 polymorph structure.

Combined with previous *ab initio* results and supported by experimental findings, our theoretical investigations show that tuning the energy difference of competing B1 vs. B4 structures is a viable approach to control the inherent toughness of B1 transition-metal-Al-N solid solutions. More generally, present AIMD simulations emphasize the importance of exploiting phase metastability as a trigger for activating transformation toughening and plastic deformation in materials at extreme mechanical-loading conditions.

**Acknowledgements**

All simulations were carried out using the resources provided by the Swedish National Infrastructure for Computing (SNIC), on the Clusters located at the National Supercomputer Centre (NSC) in Linköping, the Center for High Performance Computing (PDC) in Stockholm, and at the High Performance Computing Center North (HPC2N) in Umeå, Sweden. We gratefully acknowledge financial support from the Competence Center Functional Nanoscale Materials (FunMat-II) (Vinnova grant no 2016–05156), the Swedish Research Council (VR) through Grant No. 2019-05600, the Swedish Government Strategic Research Area in Materials Science on Functional Materials at Linköping University (Faculty Grant SFO-Mat-LiU No. 2009-00971), and the Knut and Alice Wallenberg Foundation through Wallenberg Scholar project (Grant No. 2018.01941). D.G.S. gratefully acknowledges financial support from the Olle Engkvist Foundation.

**Tables**

|  | TiN | Ti$_{0.75}$Al$_{0.25}$N | Ti$_{0.5}$Al$_{0.5}$N | Ti$_{0.25}$Al$_{0.75}$N |
|---|---|---|---|---|
| $C_{11}$ (GPa) | 650±50 | 592±38 | 552±28 | 528±38 |
| $C_{12}$ (GPa) | 128±6 | 153±7 | 159±3 | 174±8 |
| $B$ (GPa) | 302±17 | 300±13 | 290±9 | 292±13 |

**Table I**. Room-temperature $C_{11}$, $C_{12}$ elastic constants and bulk moduli $B$ of B1 Ti$_{1-x}$Al$_x$N (0 ≤ x ≤ 0.75) solid solutions obtained from the elastic mechanical-response regime determined during AIMD tensile elongation.

| Tensile strain direction | TiN | Ti$_{0.75}$Al$_{0.25}$N | Ti$_{0.5}$Al$_{0.5}$N | Ti$_{0.25}$Al$_{0.75}$N |
|---|---|---|---|---|
| [001] | | | | |
| γ | 39 | 39 | 39 | 37 |
| U$_R$ (GPa) | 2.5 | 2.4 | 3.0 | 2.8 |
| δ$_y$ (%) | 10 | 10 | 12 | 12 |
| U$_T$ (GPa) | ≈U$_R$ | ≈U$_R$ | »U$_R$ | »U$_R$ |
| δ$_f$ (%) | 12 | 12 | ≈50 | ≈50 |
| Deformation mechanism | elastic | elastic | elastic → transformation toughening | elastic → transformation toughening |
| Failure mechanism | sudden cleavage | sudden cleavage | slow bond fraying | slow bond fraying |
| Mechanical behavior | hard/brittle | hard/brittle | hard/supertough | hard/supertough |
| [110] | | | | |
| γ | 54 | 55 | 54 | 56 |
| U$_R$ (GPa) | 5.5 | 5.4 | 4.4 | 4.7 |
| δ$_y$ (%) | 16 | 16 | 14 | 14 |
| U$_T$ (GPa) | ≈U$_R$ | ≈U$_R$ | ≈(3/2)·U$_R$ | ≈(3/2)·U$_R$ |
| δ$_f$ (%) | 18 | 18 | ≈24 | ≈24 |
| Deformation mechanism | elastic | elastic | elastic → transformation toughening | elastic → transformation toughening |
| Failure mechanism | sudden cleavage | sudden cleavage | bond fraying | bond fraying |
| Mechanical behavior | hard/brittle | hard/brittle | hard/tough | hard/tough |
| [111] | | | | |
| γ | 71 | 66 | 64 | 70 |
| U$_R$ (GPa) | 7.3 | 6.0 | 4.9 | 5.5 |
| δ$_y$ (%) | 18 | 16 | 14 | 14 |
| U$_T$ (GPa) | ≈U$_R$ | ≈U$_R$ | ≳U$_R$ | ≈(3/2)·U$_R$ |
| δ$_f$ (%) | 20 | 18 | ≈20 | ≈24 |
| Deformation mechanism | elastic | elastic | elastic | elastic → transformation toughening |
| Failure mechanism | sudden cleavage | sudden cleavage | rapid bond fraying | bond fraying |
| Mechanical behavior | hard/brittle | hard/brittle | hard/partially tough | hard/tough |

**Table II**. Mechanical properties and behavior of B1 Ti$_{1-x}$Al$_x$N (0 ≤ x ≤ 0.75) solid solutions as predicted via AIMD simulations at 300 K. The symbols represent: γ = ideal tensile strength, U$_R$ = modulus of resilience, U$_T$ = tensile toughness, δ$_y$ = yield strain, δ$_f$ = elongation at fracture.

**Figures**

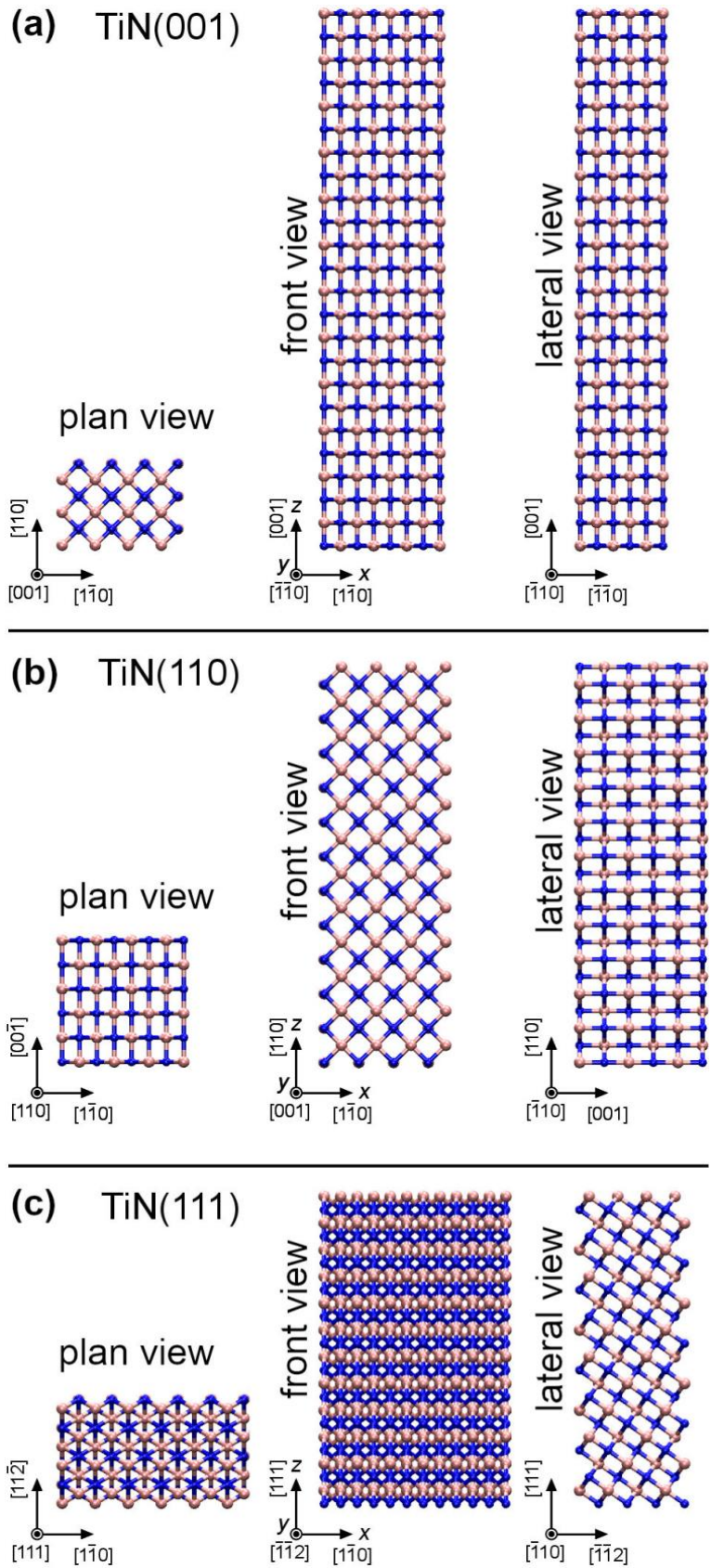

**Fig. 1**. Orthographic view of B1 supercell structures with **(a)** [001], **(b)** [110], and **(c)** [111] vertical ($z$) orientation used for AIMD tensile and shear deformation. The cation sublattice is formed of one metal species (pink spheres), while the anion sublattice is represented with blue spheres. In AIMD simulations, uniform tensile deformation is applied along vertical ($z$) directions.

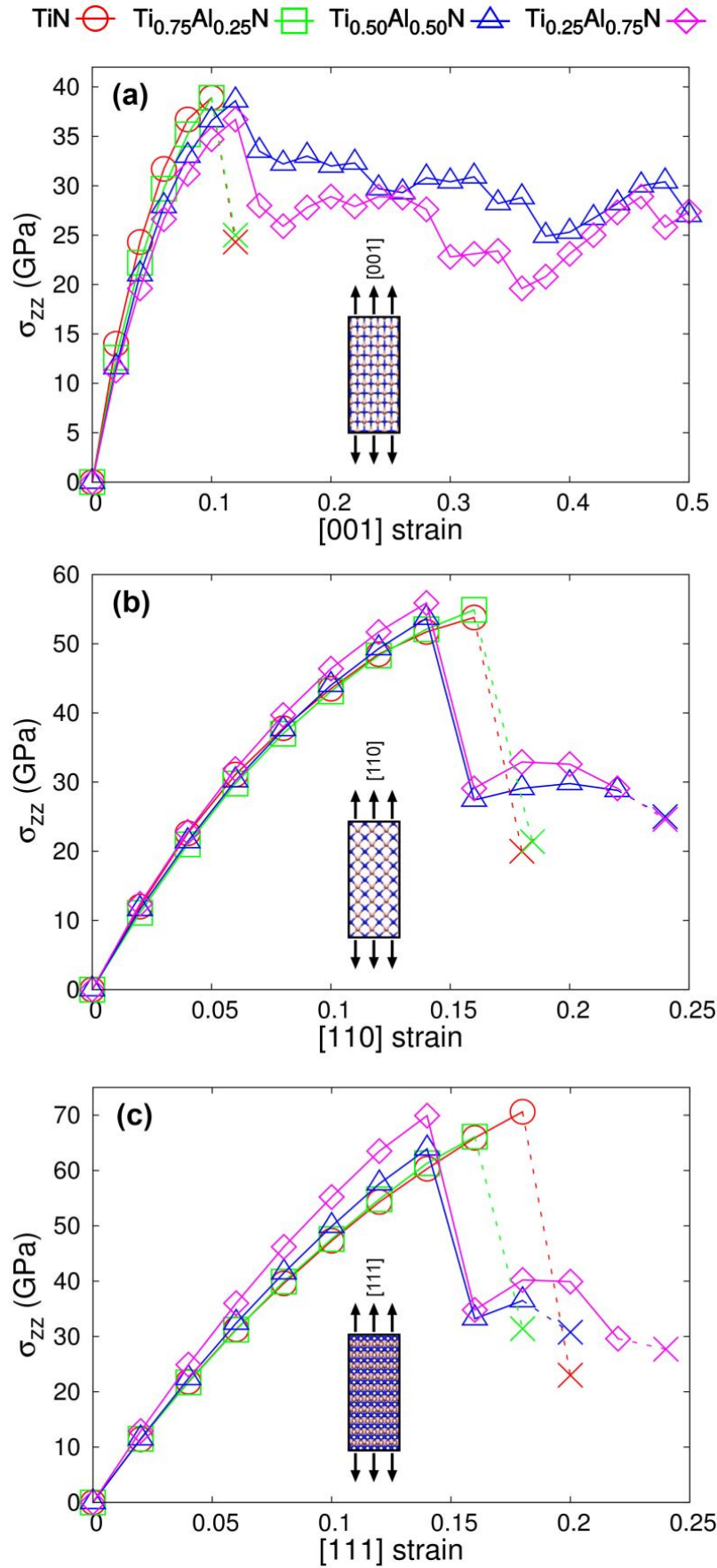

**Fig. 2**. Ti$_{1-x}$Al$_x$N(001) stress/strain curves determined via AIMD simulations at 300 K for tensile deformation along **(a)** [001], **(b)** [110], and **(c)** [111] crystallographic directions. Brittle fracture conditions are indicated by dashed curves terminated with "×" symbols. The insets are schematic representations of tensile-strained simulation supercells.

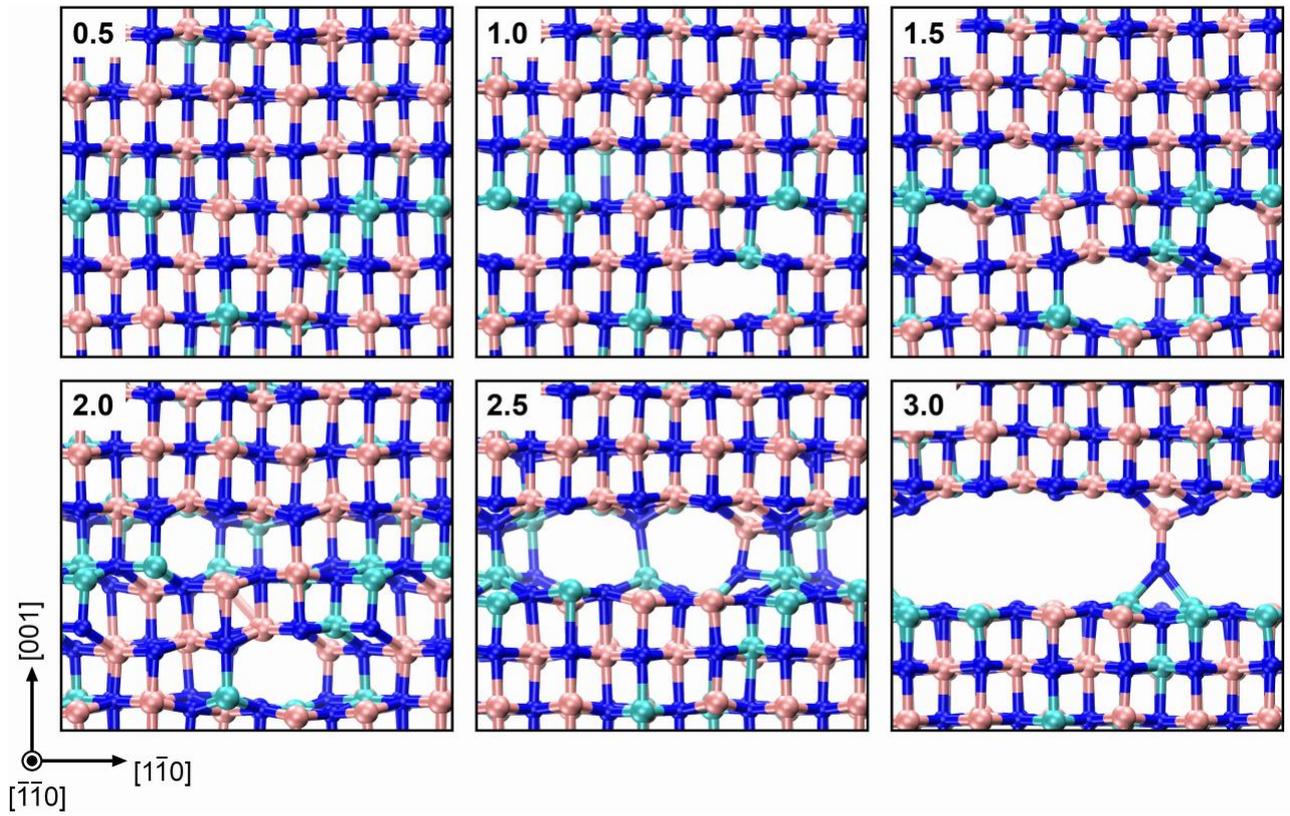

**Fig. 3**. Cleavage of B1 Ti$_{0.75}$Al$_{0.25}$N(001) on the (001) plane due to [001] tensile deformation of 12%. The AIMD simulation time passed since the alloy has been extended by 12% is in units of ps (see upper-left corners in each panel). Color legend for atomic species: blue = N, pink = Ti, cyan = Al. The dynamics bonds have cutoff lengths of 2.6 Å.

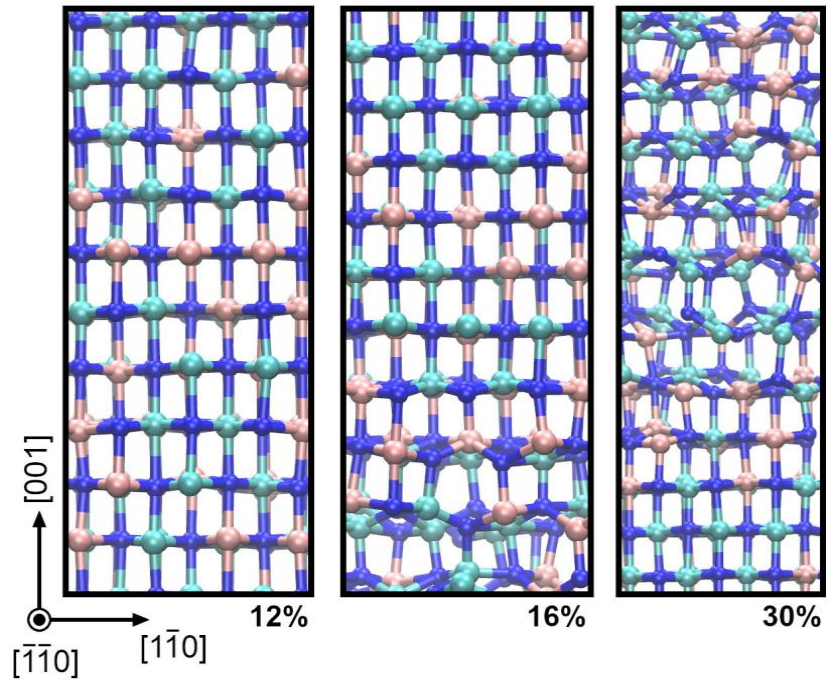

**Fig. 4**. Local B1→B4 structural transitions in tensile-strained Ti$_{0.50}$Al$_{0.50}$N(001). The three orthographic views are AIMD snapshots taken at elongations of (from left to right) 12% (which corresponds to the Ti$_{0.50}$Al$_{0.50}$N(001) yield point in **Fig. 2(a)**, 16%, and 30%. The dynamics bonds have cutoff lengths of 2.6 Å. Color legend: blue = N, pink = Ti, cyan = Al.

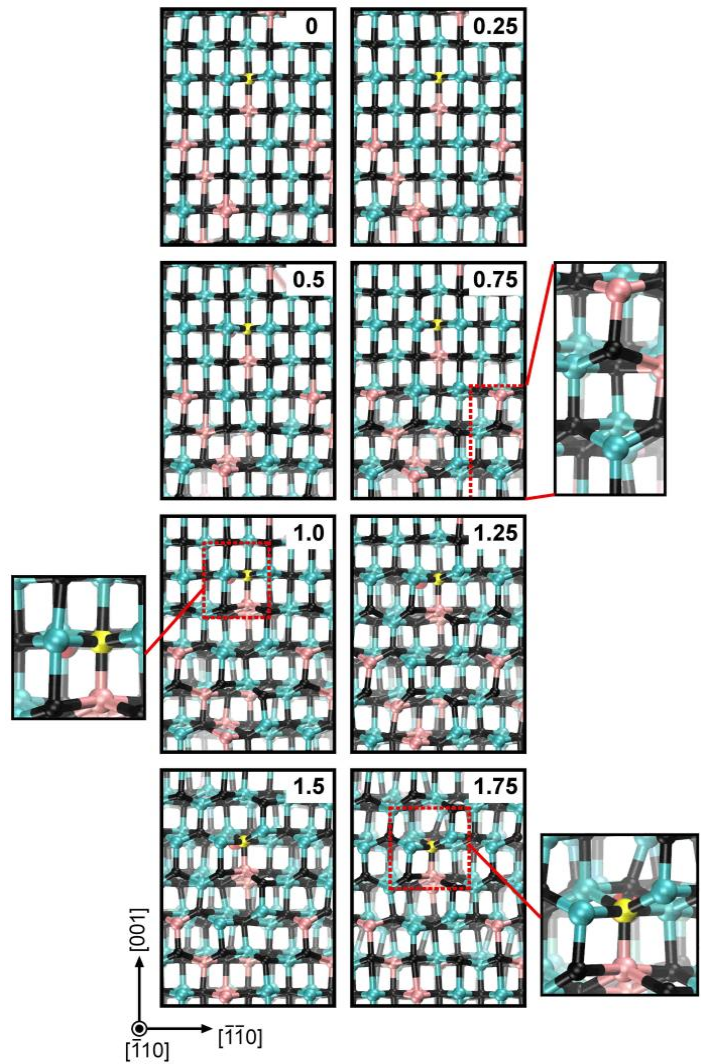

**Fig. 5**. Orthographic view of local B1→tetragonal→B4 structural transitions in Ti$_{0.25}$Al$_{0.75}$N(001) elongated by 14%. Each panel is labeled with the simulation time (ps). N atoms are colored in black, while Ti/Al atoms are pink/cyan spheres. The dynamics bonds have cutoff lengths of 2.6 Å. The magnification at 0.75 ps shows a local tetragonal (Ti,Al)N environment. The insets at 1.0 and 1.75 ps facilitate visualization of local tetragonal → B4 transformations (schematically represented in **Fig. 6**) which proceed via lattice shearing within the (001) *xy* plane: a N atom (yellow) and an Al atom (red) located on different (–110) layers progressively align on a same direction, normal to the page.

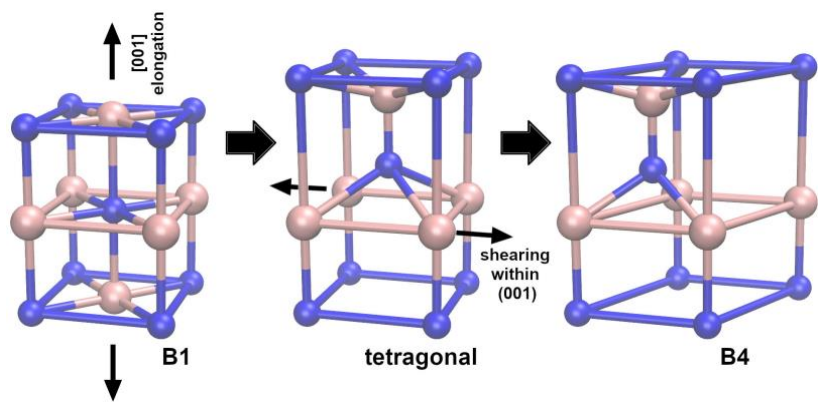

**Fig. 6**. Schematic representation of B1→tetragonal→B4 structural transitions induced by [001] tensile deformation. Spheres of different colors indicate metal and non-metal atoms.

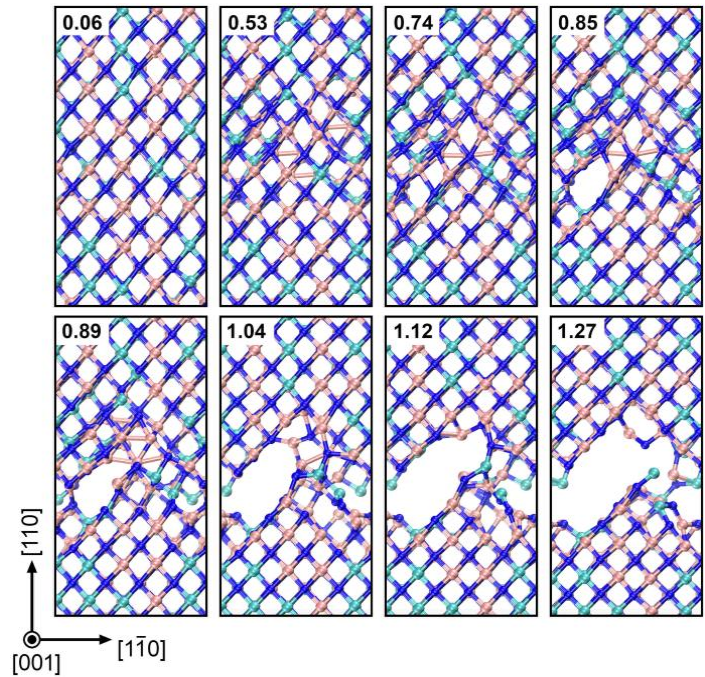

**Fig. 7**. AIMD snapshots of Ti$_{0.75}$Al$_{0.25}$N(110) brittle cleavage dynamics taken over a time-window of ≈1.3 ps at a constant [110] tensile elongation of 18% (see green curve in **Fig. 2(b)**). The time progression (ps) is indicated in the upper-left corner of each panel. Note that, although the strain is along [110], fracture develops on (001) crystallographic planes. The dynamics bonds have cutoff lengths of 2.6 Å. Color legend: blue = N, pink = Ti, cyan = Al.

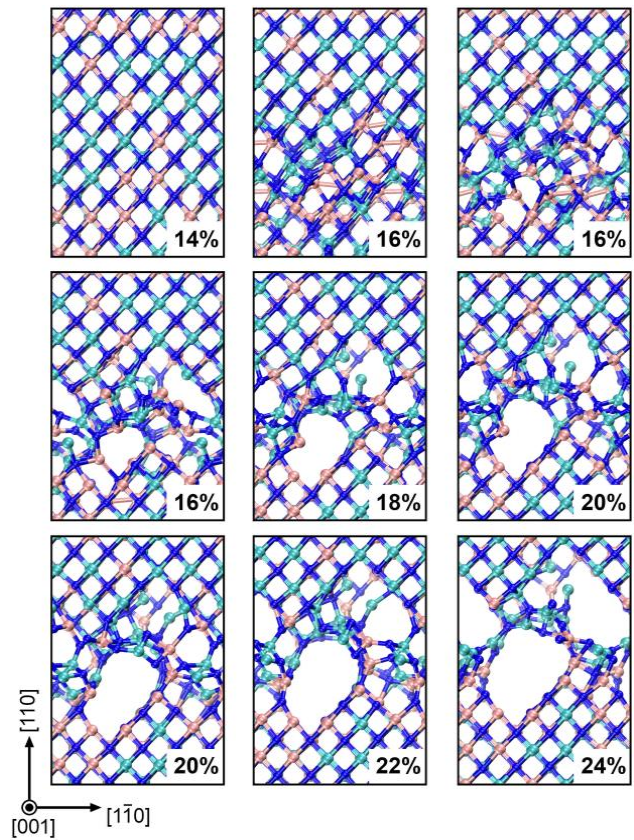

**Fig. 8**. AIMD snapshot sequence of Ti$_{0.5}$Al$_{0.5}$N(110) tensile-strained from 14% (yield point, see **Fig. 2(b)**) up to 24%. The three snapshots at 16% strain are taken at different simulation times during ≈1 ps. The alloy fractures at ≈22–24% strain. Note that open surfaces form primarily on (001) planes. The dynamics bonds have cutoff lengths of 2.6 Å. Color legend: blue = N, pink = Ti, cyan = Al.

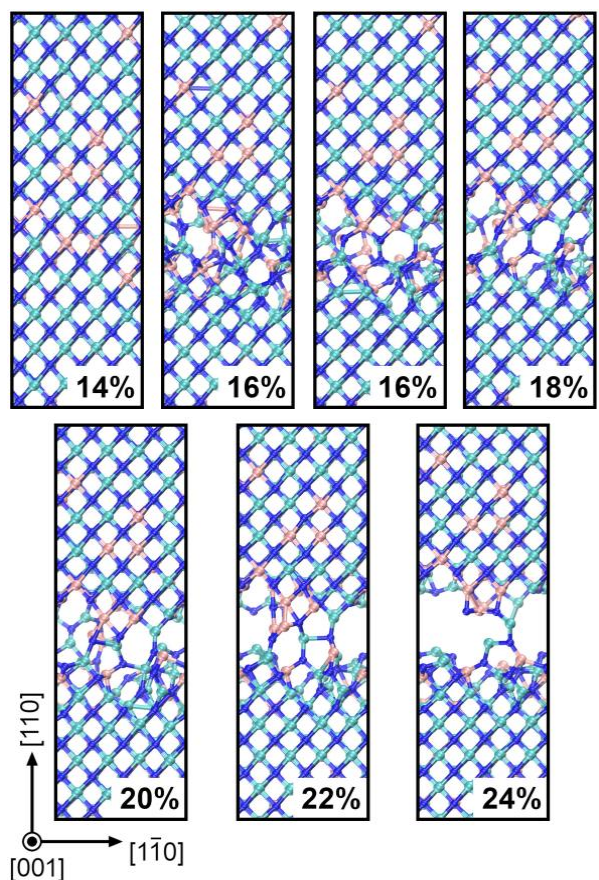

**Fig. 9**. AIMD snapshots of Ti$_{0.25}$Al$_{0.75}$N(110) subject to uniaxial strains between 14% and 24%. The two snapshots at 16% strain are taken at different simulation times. The alloy fracture occurs at ≈22–24% strain. Note that open surfaces form primarily on (001) planes. The dynamics bonds have cutoff lengths of 2.6 Å. Color legend: blue = N, pink = Ti, cyan = Al.

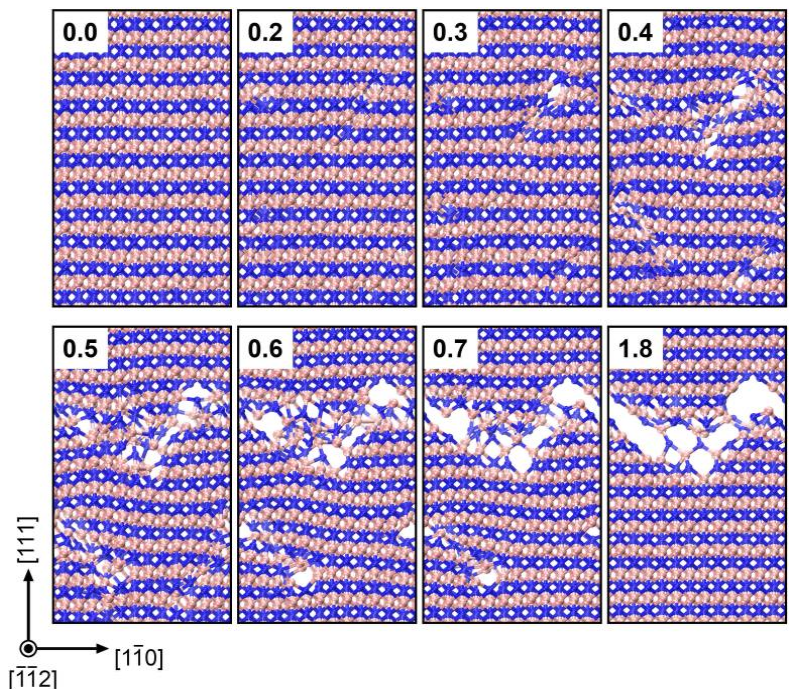

**Fig. 10**. AIMD snapshots of TiN(111) sudden breakage when maintained at a constant [111] elongation of 20%. The numbers in each panel indicate time progression (ps). The dynamics bonds have cutoff lengths of 2.6 Å. Color legend: blue = N, pink = Ti.

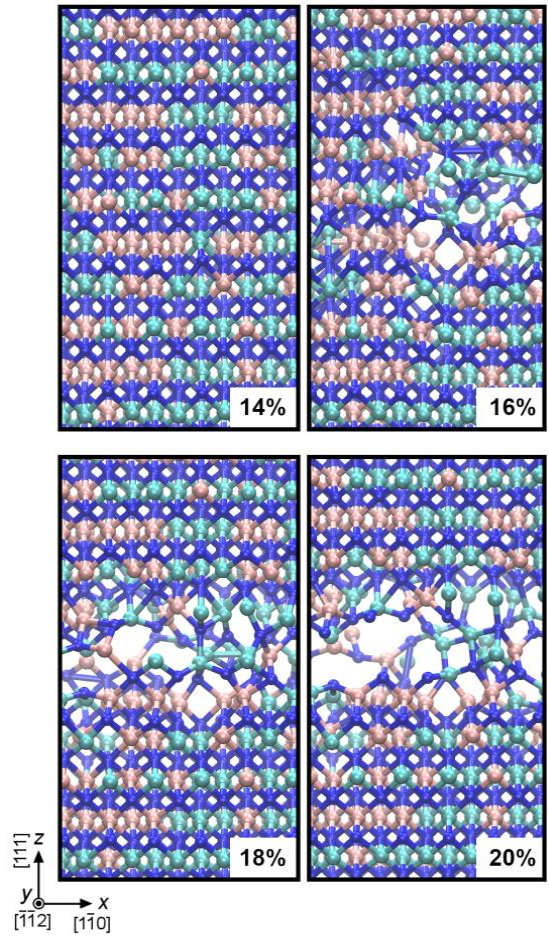

**Fig. 11.** AIMD snapshots of Ti$_{0.5}$Al$_{0.5}$N(111) breakage during [111] tensile deformation. The dynamics bonds have cutoff lengths of 2.6 Å. Color legend: blue = N, pink = Ti, cyan = Al.

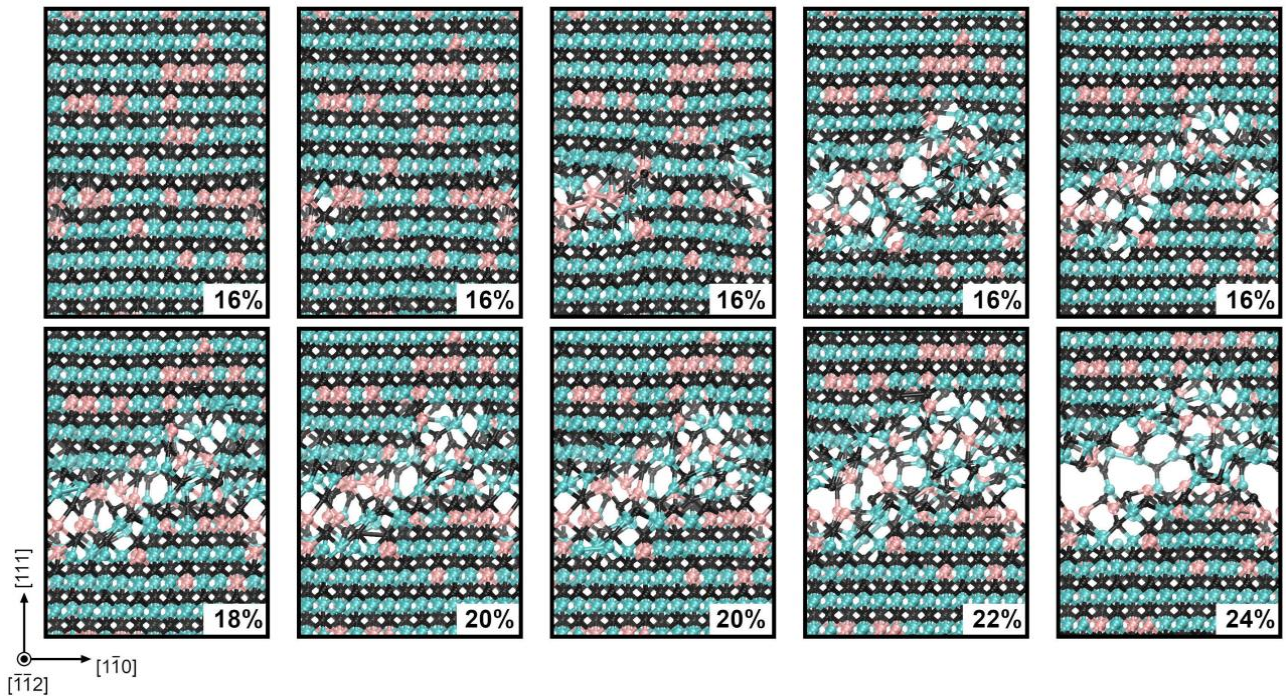

**Fig. 12.** AIMD snapshots of Ti$_{0.25}$Al$_{0.75}$N(111) breakage during [111] tensile deformation. The upper panels (16% deformation) show atomic configurations at different simulation times, during a timeframe of ≈1.5 ps. N atoms are colored in black, while Ti/Al atoms are pink/cyan spheres. The dynamics bonds have cutoff lengths of 2.6 Å.